\documentclass[preprint]{ptephy_v1}

\preprintnumber{XXXX-XXXX} 
\usepackage{hyperref}
\hypersetup{
    breaklinks,
    unicode,
    linktoc=all,
    bookmarksnumbered=true,
    bookmarksopen=true,
    colorlinks,
    linkcolor=blue,
    citecolor=blue,
    urlcolor=blue,
    plainpages=false,
    pdfstartview=FitH,
    pdfborder={0 0 0},
    linktocpage
}

\newcommand{\ts}{\textsuperscript}

\begin{document}

\title{Deformation from zinc to zirconium}


\author{Sidong Chen}
\affil{School of Physics, Engineering and Technology, University of York, Heslington, York, YO10 5DD, UK  \email{sidong.chen@york.ac.uk}}

\author{Frank Browne}
\affil{School of Physics and Astronomy, The University of Manchester, Manchester M13 9PL, United Kingdom \email{frank.browne@manchester.ac.uk}}

\author{Tom\'as R. Rodr\'iguez}
\affil{Grupo de F\'isica Nuclear, Dpto. EMFTEL and IPARCOS, Universidad Complutense de Madrid, 28040, Madrid, Spain \email{tomasrro@ucm.es}}

\author[4,5]{Volker Werner\thanks{These authors contributed equally to this work}}
\affil[4]{Technische Universität Darmstadt, Institut für Kernphysik, D-64289 Darmstadt, Germany}
\affil[5]{Helmholtz Forschungsakademie Hessen für FAIR (HFHF), D-64289 Darmstadt, Germany \email{vw@ikp.tu-darmstadt.de}}


\begin{abstract}%
Extensive gamma-ray spectroscopy of very neutron-rich nuclei of isotopes between the Ni and Sn isotopic chains was facilitated by the high luminosity LH2 target system, MINOS. Results show a persistence of deformation when going beyond the $N=60$ threshold of the transition between spherical to deformed ground states at $N<60$ and $N>60$, respectively. Close to \ts{78}Ni, a more detailed image of the $N>50$ Zn isotopes shows an erosion of the $N=50$ shell closure, with core-breaking effects needed from theoretical models to replicate observation. As well as the experimental results indicating collective effects, the projected generator coordinate method is discussed in detail within the context of the neutron-rich Ge isotopes. 
\end{abstract}

\subjectindex{xxxx, xxx}

\maketitle

\section{Introduction}
The neutron-rich mid-shell region between the Ni and Sn isotopic chains is characterised by a transition from sphericity at the $N=50$ isotopes towards more collective structures. The nature and effects of this transition, however, change dramatically when moving from $Z=30$ to $Z=40$. 
Whereas coexisting shapes may already be found in $^{78}$Ni \cite{Taniuchi78Ni}, the Zn isotopes just north-east of the doubly magic \ts{78}Ni are particularly sensitive to the effects of a weakened $N=50$ shell closure~\cite{shand_plb_2017}. Moreover, the Ge and Se isotopes have been shown to exhibit rather different pathways to deformation from their semi-magic members \cite{PRC_96_011301_2017,PRC_95_041302_2017,Lizarazo_2020_PRL,Gerst_2022_PRC}. Theories tend to produce shape co-existence, e.g., between prolate and oblate structures in Se isotopes around $N=56$, and give rise to triaxiality, from soft to more rigid structures, in Ge
isotopes at $N=52,54$. Experimentally, hints for both have been identified from recent investigations at the RIBF. The decay behaviour of isomeric states in Se isotopes~\cite{Lizarazo_2020_PRL} 
may be connected to co-existing shapes, similarly to Kr isotopes, and the energy ordering and decay branches of the $2^+_1$, $2^+_2$ , and $3^+_1$ states in Ge isotopes
hints to a degree of triaxiality.

While, up to Kr, the onset of collectivity into the open neutron shell was found to be rather smooth, the situation changes drastically in the Sr and Zr isotopes, as an effect of the neutron $d_{5/2}$ and $s_{1/2}$ closures at $N=56,58$, respectively. This effect is strongest in the Zr isotopes with a near doubly-magic behaviour of $2^+_1$ energies at $^{96}$Zr, and a sharp drop toward deformation at $N=100$. These shape (phase) transitional features at $Z \approx 40$ have been subject of much theoretical efforts, see, e.g., Refs.~\cite{Heyde_2011_RMP,Sieja_2009_PRC,Sarriguren_2015_PRC,PRC_90_034306_2014,Togashi_2016_PRL}, and more recently been put into perspective by microscopic calculations \cite{Togashi_2016_PRL} as so-called type-II shape evolution, and by collective model calculations \cite{Gavrielov2019} as so-called intertwined quantum phase transitions. Either of the latter approaches identify the coexistence of multiple distinctly deformed structures, including spherical, prolate, oblate and triaxial, although with some variations among the models, which started to be tested by experiments, most importantly at the shape-phase transitional $^{98}$Zr~\cite{Witt2018,Singh2018,Karayonchev2020}. 

The coexistence of multiple deformations in one isotope, and phase- or change-over transitions between them over chains of isotopes, must be looked at and understood in a common framework and needs data on the neutron-rich isotopes toward and beyond $N=60$. In particular, experimental tests of a potential shell gap at $^{110}$Zr~\cite{Paul_2017_PRL}, amidst the most deformed Zr isotopes, has been an important aspect of the experimental program reported here. Closer to the \ts{78}Ni isotopes, how collectivity arises in the Zn isotopes beyond $N=50$ assists in the understanding of the origin of the unexpected shape coexisting phenomenon inferred in \ts{78}Ni.

\section{Projected generator coordinate method calculations}
A well-suited theoretical framework to study intrinsic deformations from a microscopic viewpoint is the projected generator coordinate method (PGCM) [also called symmetry-conserving configuration mixing method (SCCM) or projected configuration mixing method (PCM)]. This method is based on the variational principle and the nuclear states with total angular momentum, $J$, its third component, $M$, and parity, $\pi$, are defined as linear combinations of symmetry-restored mean-field-like wave functions (see Refs.~\cite{PRC_78_024309_2008,PRC_81_064323_2010,JPG_46_013001_2019} and references therein):
\begin{equation}
\label{eq:pgcm_wf}
|JM\pi;\sigma\rangle=\sum_{q}\sum_{K=-J}^{J}f^{J\pi,\sigma}_{q,K}|JMK\pi;q\rangle.
\end{equation}  
Here, the projected wave functions, $|JMK\pi;q\rangle$, are obtained by the symmetry restoration of intrinsic Hartree-Fock-Bogoliubov (HFB) vacuum states:
\begin{equation}
|JMK\pi;q\rangle=P^{J}_{MK}P^{N}P^{Z}P^{\pi}|\Phi_{q}\rangle
\label{eq:proj_wf}
\end{equation}
where $P^{J}_{MK}$, $P^{N}$, $P^{Z}$, and $P^{\pi}$ are the projection operators onto good angular momentum ($K$ is the component of the angular momentum vector on the $z$-axis of the body-frame), number of neutrons and protons, and parity, respectively~\cite{RingSchuck}. The Bogoliubov vacua ($|\Phi_{q}\rangle$) are obtained by solving plain or projected HFB equations with constraints to the generating coordinates, $\lbrace q\rbrace$. Finally, the coefficients of the linear combination in Eq.~\ref{eq:pgcm_wf}, hence, the nuclear wave functions and the nuclear energies, are found by minimizing the PGCM energy that leads to the Hill-Wheeler-Griffin (HWG) equations:
\begin{equation}
\sum_{q',K'}\left(\mathcal{H}^{J\pi}_{q,K;q',K'}-E^{J\pi;\sigma}\mathcal{N}^{J\pi}_{q,K;q',K'}\right)f^{J\pi,\sigma}_{q',K'}=0.
\label{eq:hwg}
\end{equation}
The HWG equations are generalized eigenvalue problems that are solved by defining a set of orthonormal states from the original set of non-orthogonal (and, possible, linear-dependent) states given by Eq.~\ref{eq:proj_wf}. Hence, the first step is the diagonalization of the norm overlap matrix:
\begin{equation}
\sum_{q',K'}\mathcal{N}^{J\pi}_{q,K;q',K'}u^{J\pi}_{q',K';\Lambda_{i}}=n^{J\pi}_{\Lambda_{i}}u^{J\pi}_{q,K;\Lambda_{i}}
\label{eq:norm_eigen}
\end{equation}
Then, the orthonormal states that define the so-called \textit{natural basis} are obtained as:
\begin{equation}
|\Lambda^{J\pi}_{i}\rangle=\sum_{q,K}\frac{u^{J\pi}_{q,K;\Lambda_{i}}}{\sqrt{n^{J\pi}_{\Lambda_{i}}}}|JMK\pi;q\rangle
\label{eq:nat_basis}
\end{equation}
In the above expression the states with very small eigenvalues of the norm overlap matrix, $n^{J\pi}_{\Lambda_{i}}<\epsilon$, are discarded from the definition of the natural basis to remove linear dependencies and numerical inconsistencies. Finally, the PGCM wave functions, Eq.~\ref{eq:pgcm_wf}, are now expressed in terms of the natural basis:
\begin{equation}
|JM\pi;\sigma\rangle=\sum_{i}G^{J\pi;\sigma}_{\Lambda_{i}}|\Lambda^{J\pi}_{i}\rangle.
\label{eq:pgcm_wf_nat}
\end{equation}
The HWG equations (Eq.~\ref{eq:hwg}) are then transformed into regular eigenvalue problems:
\begin{equation}
\sum_{j}\langle\Lambda^{J\pi}_{i}|\hat{H}|\Lambda^{J\pi}_{j}\rangle G^{J\pi;\sigma}_{\Lambda_{j}}=E^{J\pi;\sigma}G^{J\pi;\sigma}_{\Lambda_{i}}
\end{equation}

In the present implementation of the PGCM method, performed with the Gogny D1S functional~\cite{NPA_428_23_1984}: \textbf{a)} the intrinsic states are obtained with the variation after particle number projection (VAPNP) method (instead of HFB) to incorporate efficiently pairing correlations within such intrinsic states~\cite{NPA_696_467_2001}; \textbf{b)} intrinsic quadrupole deformations, parametrized by $(\beta_{2},\gamma)$, are explicitly explored as collective generating coordinates, $\lbrace q\rbrace$; \textbf{c)} parity-symmetry is not broken, therefore, only positive parity states can be studied and parity projection is not needed; and, \textbf{d)} no explicit multi-quasiparticle nor time-reversal symmetry breaking (such as cranking) HFB-like states are considered. This favours the variational exploration of the ground state with respect to the excited states and some stretching of the spectrum is expected. Because all the particles are taking part of the system, intrinsic shapes, given by the properties of the underlying HFB-like states, are well-defined theoretical quantities. The relevance of the different shapes in the structure of a given nucleus can be studied qualitatively by the behaviour of the VAPNP energy as a function of the collective deformations, $(\beta_{2},\gamma)$. These are the so-called total energy surfaces (TES). A more quantitative information about the shape of the nucleus is provided by the collective wave function (c.w.f.) of the individual PGCM states. These quantities represent the weight of the different intrinsic shape in the building of each PGCM wave function along the $(\beta_{2},\gamma)$ plane and they are normalized to 1. The explicit expression for these quantities in terms of the coefficients of the PGCM wave functions (Eq.~\ref{eq:pgcm_wf_nat}) is given by:
\begin{equation}
|F^{J\pi;\sigma}(\beta_{2},\gamma)|^{2}=\left|\sum_{i,K}G^{J\pi;\sigma}_{\Lambda_{i}}u^{J\pi}_{(\beta_{2},\gamma),K}\right|^{2}
\label{eq:cwf_def}
\end{equation} 
\begin{figure}[t]
\centering\includegraphics[width=1.\textwidth]{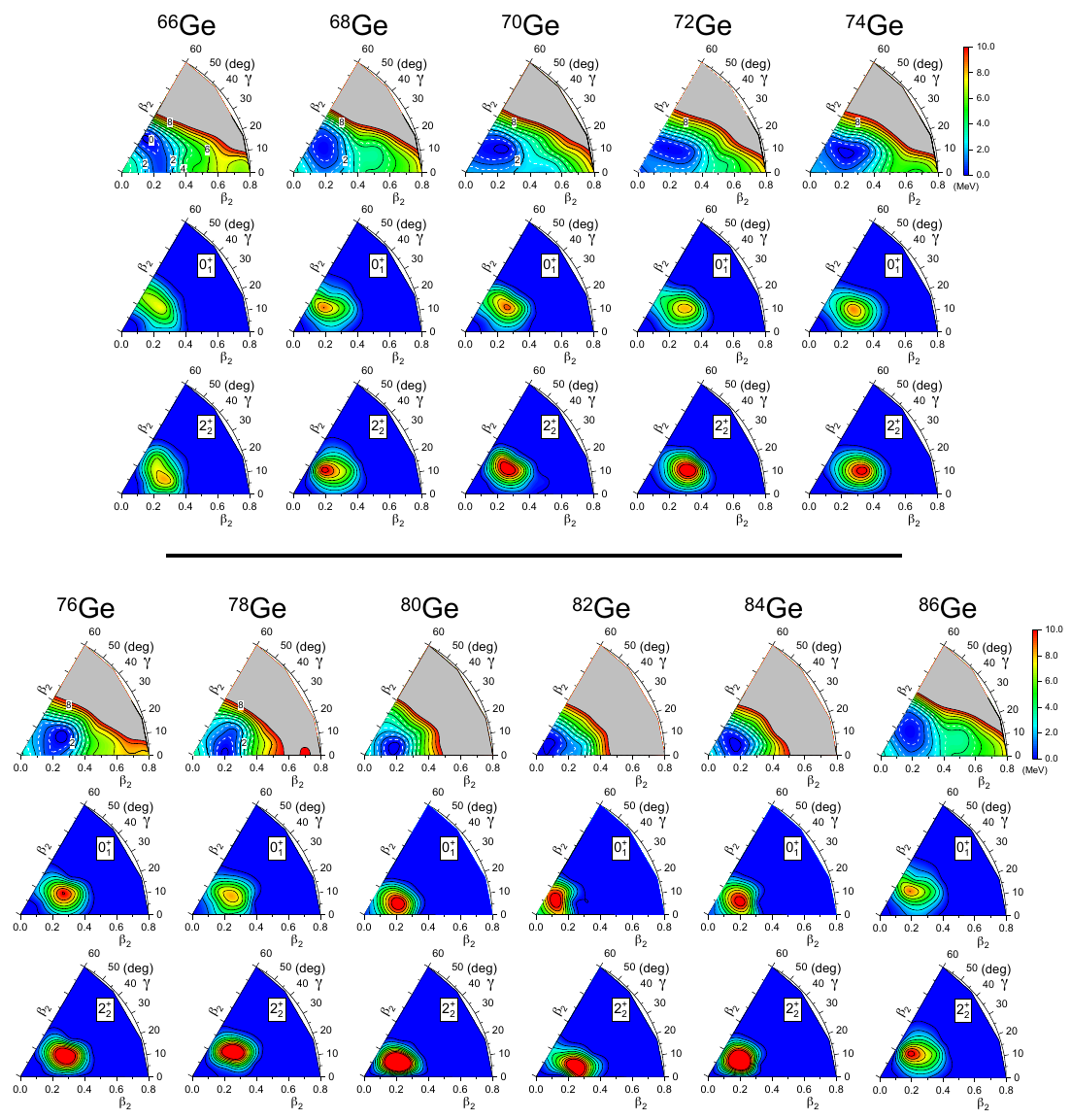}
\caption{From the top to the bottom row, VAPNP total energy surfaces (TES), and collective wave functions for ground state, and $\gamma$-band head states as a function of $(\beta_{2},\gamma)$ quadrupole deformations are shown. The top (bottom) panel corresponds to $^{66-74}$Ge ($^{76-86}$Ge) isotopes.}
\label{fig:Ge_TES_CWF}
\end{figure}

As an example of application of the PGCM method to the study of shape evolution/mixing/coexistence, Ge isotopes have been computed. More applications of the present framework will be discussed in subsequent sections. A first insight into the shape evolution from neutron-deficient to neutron-rich Ge isotopes is represented by the VAPNP-TES calculated for $^{66-86}$Ge and plotted in Fig.~\ref{fig:Ge_TES_CWF}. Hence, $^{66}$Ge shows a $\gamma$-soft TES with lower oblate configurations that evolves towards more purely triaxial configurations in $^{70-76}$Ge. Approaching to the $N=50$ shell closure, where the semi-magic $^{82}$Ge nucleus shows an almost spherical minimum, the isotopes $^{78-80}$Ge present more prolate configurations. Finally, the TES of the nuclei $^{84,86}$Ge again develop triaxial deformation. Contrary to other isotopic chains in the region (see, e.g., Refs~\cite{PRC_95_041302_2017,PRC_90_034306_2014}), none of the TES shows secondary minima close in energy to the lowest minimum (a hint for shape-coexistence). The main feature of this isotopic chain is the predominant role of triaxiality, as it was already reported for neutron rich $^{84-88}$Ge~\cite{PRC_96_011301_2017}.

Once the symmetry restoration and mixing are performed, triaxially-deformed ground state (g.s.) bands ($\Delta J=2$) and $\gamma$-bands ($\Delta J=1$) associated to the g.s. bands are identified in the whole isotopic chain. The c.w.f.'s for the band-heads, $0^{+}_{1}$ and $2^{+}_{2}$ respectively, are represented in Fig.~\ref{fig:Ge_TES_CWF}. The first noticeable result is the erosion of the $N=50$ shell closure since the $0^{+}_{1}$ state do not have the maximum at the spherical point but at a small but finite value of deformation $\beta_{2}$. Decreasing the number of neutrons from $^{82}$Ge, a smooth increase of the $\beta_{2}$-deformation is observed, reaching the maximum at $^{76-74}$Ge, and, from then on, a smooth evolution towards more oblate-like and less $\beta_{2}$-deformed ground states for $^{72-66}$Ge. Above $N=50$, an increase of the $\beta_{2}$-deformation with triaxiality is also observed in the c.w.f. of $^{84,86}$Ge. A similar trend is obtained of the band-head of the $\gamma$-band, i.e., the $2^{+}_{2}$ c.w.f.'s are very much alike the g.s. counterparts (with less mixing) except for $^{66}$Ge and $^{82}$Ge where more prolate-like structures are found.     
\begin{figure}[t]
\centering\includegraphics[width=0.9\textwidth]{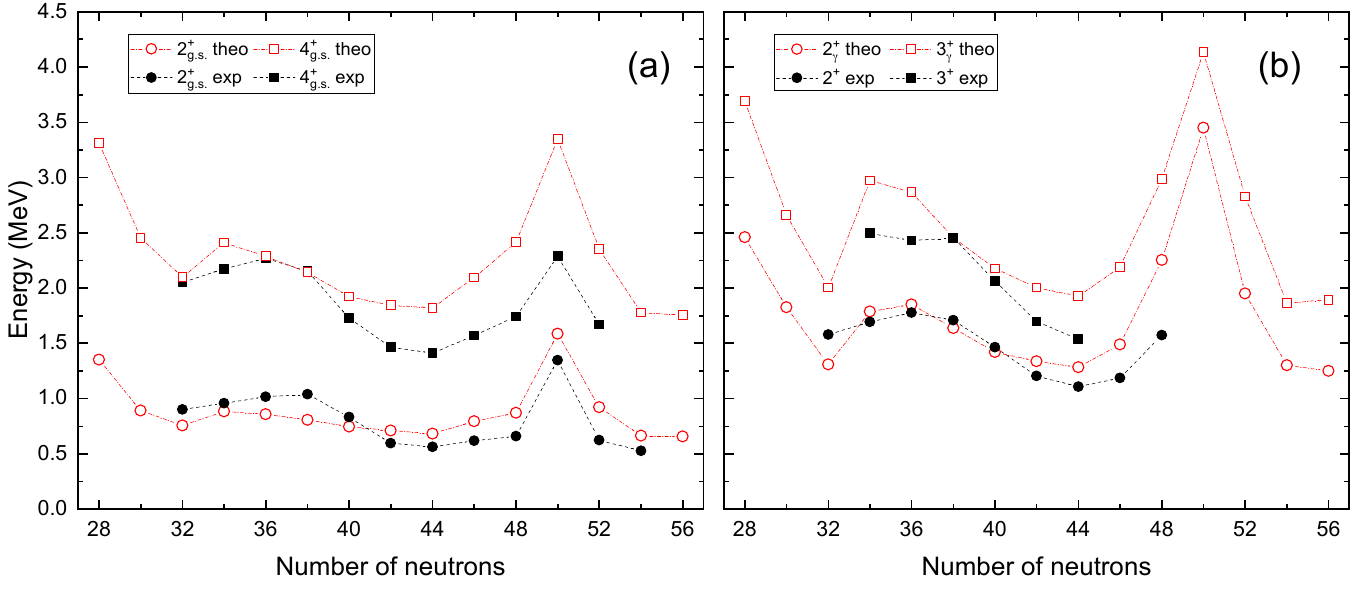}
\caption{Comparison between the PGCM and experimental~\cite{NNDC_Database} excitation energies for (a) $2^{+}$ and $4^{+}$ ground state bands, and (b) $2^{+}$ and $3^{+}$ of $\gamma$-bands in Ge ($Z=32$) isotopes.}
\label{fig:Ge_energies}
\end{figure}

The theoretical ground state band $2^{+}$, $4^{+}$ energies are shown in Fig.~\ref{fig:Ge_energies}(a) extended from $^{60-88}$Ge. As expected from the fact that they belong to the same collective band, the calculations predict a similar behaviour of the $2^{+}$ and $4^{+}$ excitation energies, although the trends are more amplified for the $4^{+}$ states. Hence, the peaks corresponding to the $N=28$ and $N=50$ shell closures are clearly visible. Then, a relatively sharp decrease of the energy from $N=28$ to $N=30$ and $32$ is observed. After a small increase from $N=32$ to $N=34$, the PGCM results predict a smooth and continuous decrease of the $2^{+}$ and $4^{+}$ excitation energies until the minimum at $^{76}$Ge is reached. A smooth increase is then obtained in $N=46-48$, and, after the peak at $N=50$, a decrease in energy is found. This evolution of the g.s.- and $\gamma$- band excitation energies predicted by the PGCM calculations is consistent with the previous analysis of the c.w.f.'s., i.e., smaller (larger) excitation energies correspond to larger (smaller) quadrupole deformation in the $0^{+}_{\text{g.s.}}$ state. The available experimental data is also plotted in Fig.~\ref{fig:Ge_energies}(a). A rather good qualitative agreement is obtained with the present PGCM calculations, although an overall overestimation of the theoretical energies is found, as expected from the absence of time-reversal symmetry breaking states in the present PGCM implementation~\cite{PLB_746_341_2015}. This qualitative agreement (except for the stretching) is better for $^{74-86}$Ge than in lighter isotopes, where the $2^{+}$ excitation energies are even below the experimental values. This result could indicate a slight overestimation of the experimental deformation for $^{64-70}$Ge isotopes in these calculations.
The appearance of $\gamma$-bands is considered as a hint of triaxial deformation in the nucleus. In Fig.~\ref{fig:Ge_energies}(b) the excitation energies for the band-head $2^{+}$ states and the connected $3^{+}$ states are shown. A similar behaviour to the ground-state band excitation energies is obtained both in the experimental data and the theoretical predictions. The very good agreement with the experimental data supports the interpretation of these nuclei as triaxially deformed.

\section{Experimental results}
The experimental results reported here are products of the first and second SEASTAR campaigns~\cite{doornenbal_rapr_2015,doornenbal_rapr_2016}.
The SEASTAR (Shell Evolution And Search for Two-plus energies At RIBF) project aimed at a systematic study for $2^+$ states in the wide range of the neutron-rich nuclei accessible with the RIBF's high-intensity beams.
Central to these experiments were the DALI2+ $\gamma$-ray spectrometer~\cite{takeuchi_nima_2014} and the MINOS (MagIc Numbers Off Stability) device~\cite{obertelli_epja_2014}. MINOS comprised for both campaigns a 100-mm-long liquid-hydrogen target surrounded by a TPC to track protons originating from the target material and those knocked out from incoming beams. The 186 NaI(Tl) crystals of DALI2 (Detector Array for Low Intensity radiation) surrounded MINOS in a geometry that optimised efficiency as well as reducing the opening angle subtended to the target of MINOS to reduce Doppler broadening effects. The MINOS-DALI2 ensemble were situated between the BigRIPS and ZeroDegree spectrometers~\cite{kubo_ptep_2012} which identified incoming and outgoing beams, respectively. 

In addition to the in-beam $\gamma$-ray spectroscopy apparatus described above, decay spectroscopy was conducted at the final focal plane of the ZeroDegree spectrometer with the EURICA (Euroball-RIKEN Cluster Array) configuration~\cite{soderstrom_nimb_2013}. This array, comprising 12 clusters of seven tapered, hexagonal high-purity germanium crystals surrounded a passive beam stopper was able to tag on isomeric decays of stopped nuclei. This capability allowed for investigation of states populating isomeric states, as discussed later. 

\begin{figure}[t]
\centering\includegraphics[width=0.95\textwidth]{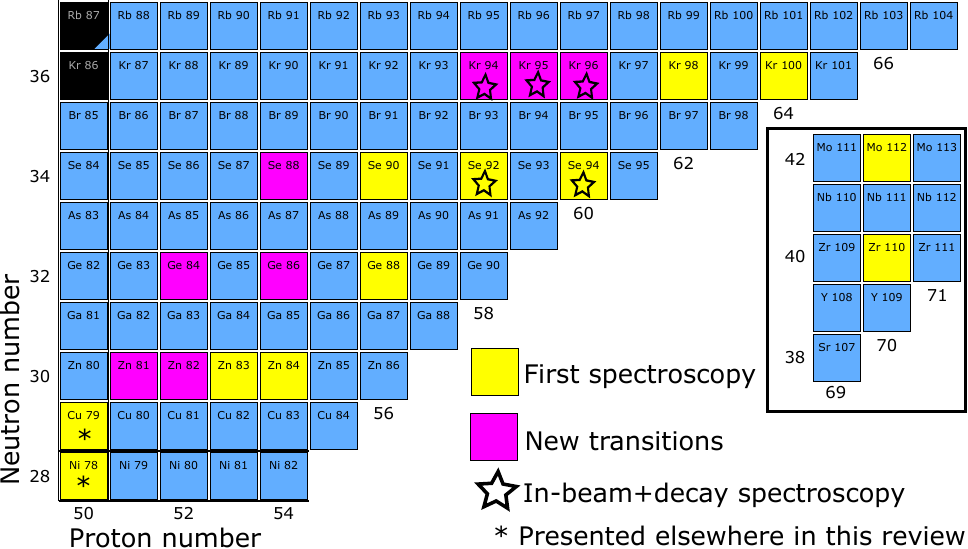}
\caption{Nuclear chart summarising the measurements discussed in this article. The portion of the chart with \ts{110}Zr and \ts{112}Mo are shown inset for aesthetic purposes. 
The \ts{78}Ni and \ts{79}Cu measurements are discussed elsewhere in this topical review. 
Measurements where the EURICA setup has been used in addition to in-beam $\gamma$-ray spectroscopy are indicated.}
\label{fig:chart}
\end{figure}

The results will be discussed in terms of isotopic chains in order of increasing proton number, that is from Zn to Zr, or going from the $Z=28$ closure towards the proton mid-shell. Figure~\ref{fig:chart} summarises the measurements discussed in this article. It highlights the broad region probed in the SEASTAR campaign, from the Zn isotopes close to the $Z=28$ and $N=50$ shell closures, to the doubly mid-shell Zr region. 

\subsection{Onset of collectivity beyond N = 50 in Zn isotopes}

\begin{figure}[t]
\centering\includegraphics[width=0.6\textwidth]{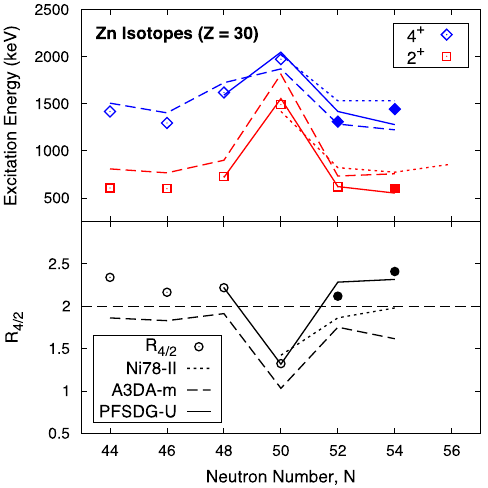}
\caption{Systematics of $E(2^+_1)$ and $E(4^+_1)$ and $R_{4/2}$ for the Zn isotopic chain. Adapted from Fig.~3 in Ref.~\cite{shand_plb_2017}. Copyright CC BY 4.0.}
\label{fig:shand_fig3}
\end{figure}

The Zn ($Z=30$) isotopes are two protons beyond the $Z=28$ magic number, with \ts{80}Zn intersecting the $N=50$ magic number. 
The low-lying level structures of \ts{80,82}Zn were investigated at the RIBF using
in-beam $\gamma$-ray spectroscopy following knockout reactions on a \ts{9}Be target~\cite{Shiga_2016_PRC}. 
Transitions from the $2^+_1$ states in \ts{80,82}Zn and the $4^+_1$ state in \ts{80}Zn were observed in DALI2 spectra.
Spectroscopy of the $2^+_1$ and $4^+_1$ states revealed behaviour consistent with that of a good $N=50$ shell closure, as shown in Fig.~\ref{fig:shand_fig3}. The second SEASTAR campaign saw the first spectroscopy of \ts{84}Zn and the first measurement of the $4^+_1$ state in \ts{82}Zn~\cite{shand_plb_2017}. These results show the expected behaviour going beyond a shell closure, that is a reduction in $E(2^+_1)$ and increase in $R_{4/2}=E(4^+_1)/E(2^+_1)$, as shown in Fig.~\ref{fig:shand_fig3}. Of the calculations employed to reproduce the results, the PFSDG-U interaction performed the best. In this interaction, breaking of the \ts{78}Ni core is allowed through the inclusion of a large neutron valence space and, unlike the A3DA-m interaction which only considers the $d_{5/2}$ orbital, includes the full $sdg$ model space. The good agreement with the models that permit core breaking suggest that the stabilising effects of the $N=50$ shell closure is localised to \ts{80}Zn, and collectivity is manifest immediately after its crossing. 

\subsection{Triaxial and prolate-oblate structures in neutron-rich Ge and Se isotopes}

\begin{figure}[t]
\centering\includegraphics[width=0.8\textwidth]{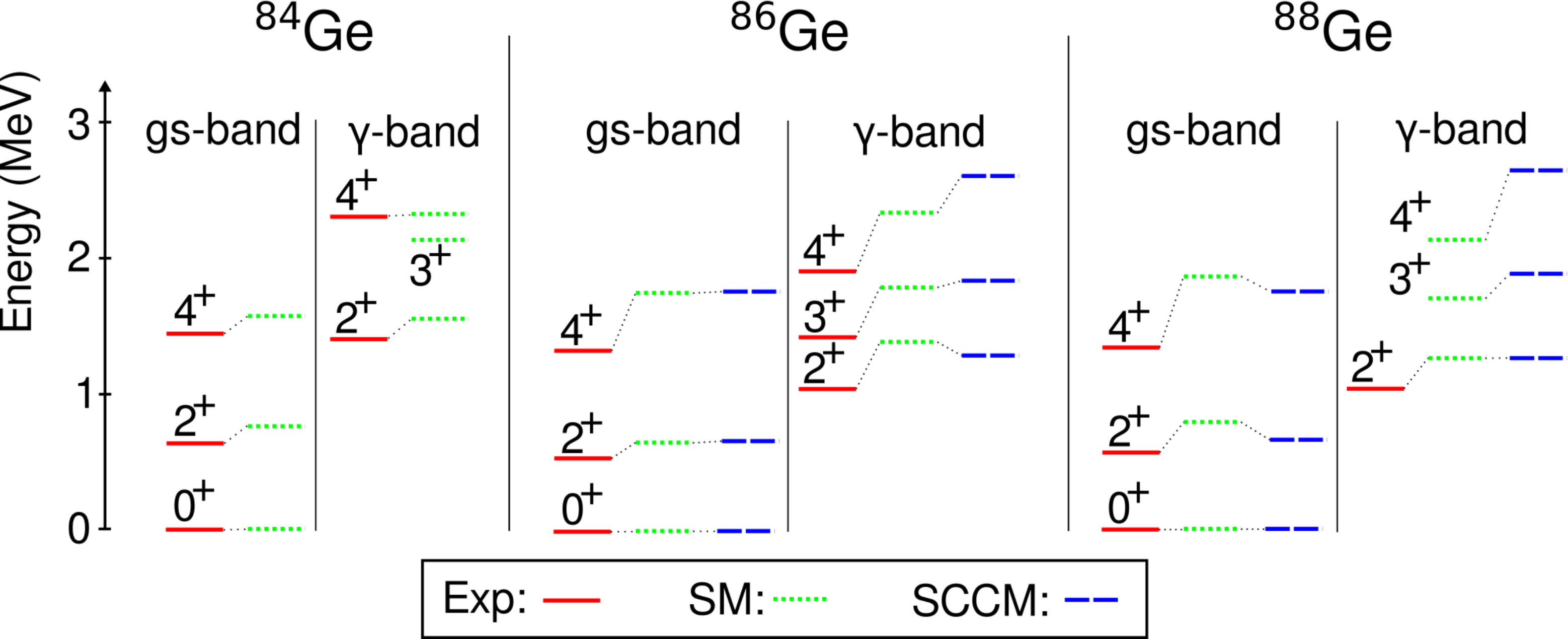}
\caption{Partial level schemes of $^{84-88}$Ge, comparing data to SCCM and shell model calculations. Figure partially adopted from Fig. 4 in Ref.~\cite{PRC_96_011301_2017}. Reprinted figure with permission from Ref.~\cite{PRC_96_011301_2017}. Copyright 2017 by the American Physical Society.}
\label{fig:gestagg}
\end{figure}

Spectroscopy of Ge and Se isotopes above the $N=50$ shell closure has given new benchmarks for the discussion of quadrupole deformation in this region between the magic shell closure and $N=58$, marking the closure of the neutron $sdg$ shell. Although the derived level schemes look rather similar, data and calculations, such as the PGCM described above, suggest different mechanisms at work. The Ge isotopes \cite{PRC_96_011301_2017} are characterized by small changes in the overall level scheme, as illustrated in Fig.~\ref{fig:Ge_energies}, with only a slight drop in $2^+_1$ energies as collectivity increases in $^{86,88}$Ge. Hovever, the level schemes indicate a drop of the $2^+_2$ energy toward $^{86}$Ge, and a hint for a potential $3^+$ state member of the odd quasi-$\gamma$ band structure, has been identified in $^{86}$Ge. Theses observations support the above-mentioned predictions of a triaxial structure developing in these isotopes, which were further supported by symmetry-conserving configuration mixing (SCCM) Gogny, as shown in Fig.~\ref{fig:Ge_TES_CWF}, and shell-model calculations. The spectroscopy resulted in level schemes akin to $^{76}$Ge \cite{Toh_76Ge} and $^{78}$Ge \cite{Forney_78Ge}, the mirror nucleus of $^{86}$Ge with respect to the $N=50$ shell, with the inverted staggering \cite{ZamCas91} of even- and odd-members of the low-lying $\gamma$ band with respect to $\gamma$-soft nuclei. Figure \ref{fig:gestagg} shows the change in level spacings within the $\gamma$ band from $^{84-88}$Ge, comparing data to SCCM and shell model calculations, with the $3^+$ state being lowered relative to the even-spin members in $^{86,88}$Ge.

\begin{figure}[tb]
\centering\includegraphics[width=0.98\textwidth]{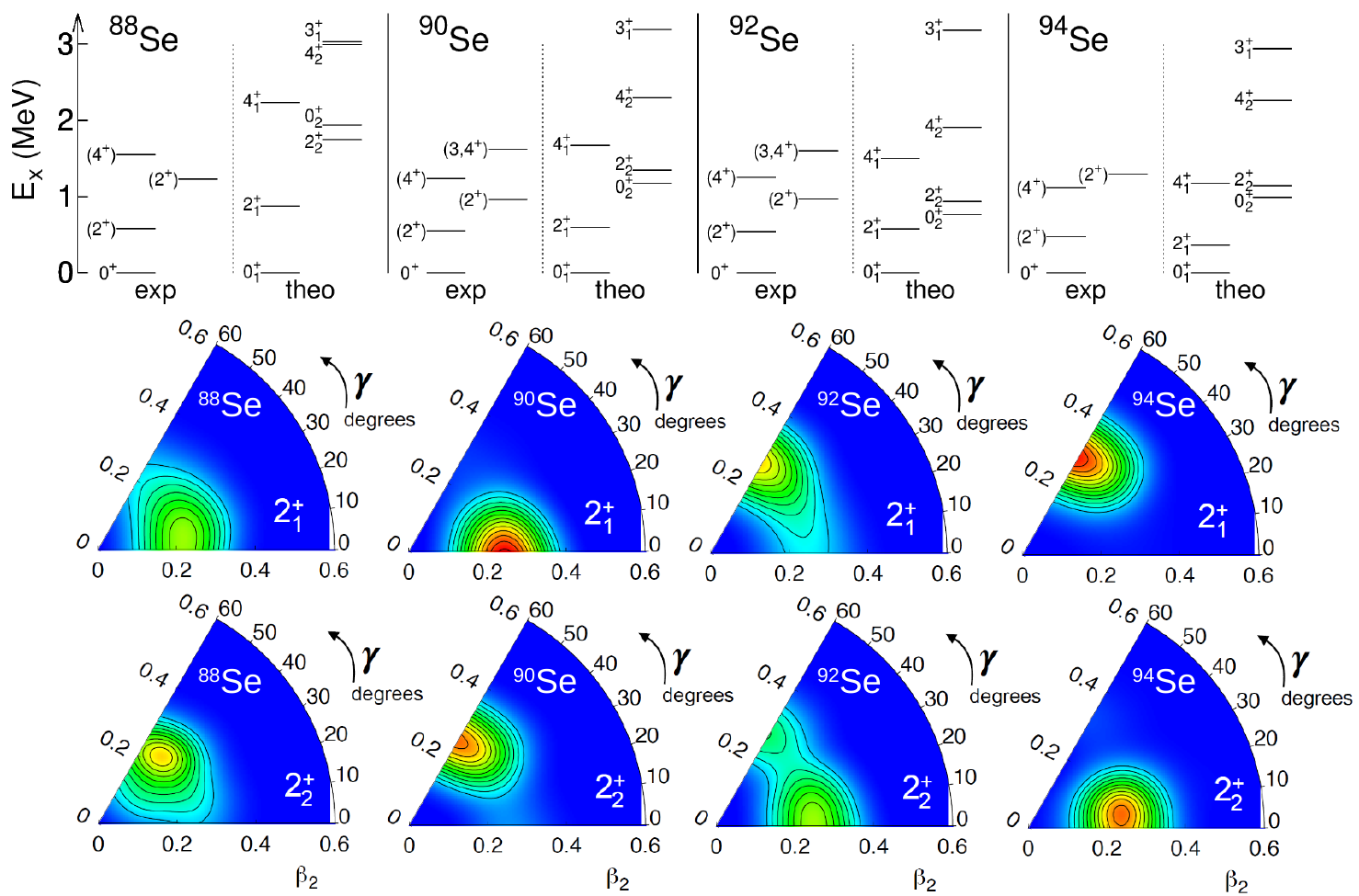}
\caption{Excitation states of Se isotopes, compared with the SCCM calculations based on the Gogny D1S interaction. The collective wave functions of the $2^+_1$ and $2^+_2$ states from the calculations are shown below the level schemes. The figure is partially adapted from Ref.~\cite{PRC_95_041302_2017}. Reprinted figure with permission from Ref.~\cite{PRC_95_041302_2017}. Copyright 2017 by the American Physical Society.}
\label{fig:Se_level_shape}
\end{figure}

For the neutron-rich Se isotopes, \ts{86}Se was the most exotic even-even isotope with a well established level scheme before this study. A possible $3^+$ level was reported in Ref.~\cite{Materna_2015_PRC}, indicating an onset of triaxial collectivity in \ts{86}Se. A similar triaxial shape was also suggested by the $B(E2)$ measurement in Ref.~\cite{Litzinger_2015_PRC}.
During the SEASTAR campaign, the $\gamma$-ray spectroscopy of neutron-rich \ts{88,90,92,94}Se was conducted following nucleon removal reactions on the liquid hydrogen target~\cite{PRC_95_041302_2017}.
The low-lying excitation level schemes were established based on $\gamma$-$\gamma$ coincidence analysis, and the spin-parity assignments based on the systematics in lighter Se isotopes and neighboring Kr isotopes.
The tentatively-assigned $2^+_1$, $4^+_1$ and $2^+_2$ states were observed in each isotope, as shown in Fig.~\ref{fig:Se_level_shape}.
The excitation states of \ts{92,94}Se were also measured by the high-resolution germanium array, EURICA~\cite{EURICA}, following the decay of their isomers in the same experiment~\cite{Lizarazo_2020_PRL}, reporting level schemes consistent with this in-beam study.
The systematics of the $E(2^+_1)$ shows a smooth, shallow drop up to $N=60$, yielding a gradual increase of collectivity.
Low-lying $2^+_2$ levels are observed below or close to the $4^+_1$ levels, indicating possible shape coexistence in these isotopes.
The experimental results were also compared calculations based on the SCCM approach with Gogny D1S effective interaction.
The calculations reproduced the excitation levels with reasonably good agreement.
The collective wave functions of the calculations suggest discrepant intrinsic deformation shapes for the $2^+_1$ and $2^+_2$ states, and a prolate-oblate shape transition through the triaxial degree of freedom in the neutron-rich Se isotopes.
The prolate-oblate coexistence and transition in the $^{92,94}$Se isotopes have also been inferred from the decay behaviour of newly-identified shape isomers in the complementary EURICA decay data \cite{Lizarazo_2020_PRL}. The search for the low-lying $0^+_2$ states and the transition probability measurements for the $2^+_1\rightarrow0^+_1$ and $2^+_2\rightarrow0^+_1$ transitions in future experiments will provide critical information for the deformed shapes in these Se isotopes.

With the occurrence and coexistence of triaxial, prolate, and oblate structures in this narrow region of the nuclear chart, a common origin likely connects all three shapes, e.g., mixing of prolate and oblate structures into resulting triaxial shapes as suggested in~\cite{CaprioIachello2004}.

\subsection{Isomer and in-beam $\gamma$-ray spectroscopy of neutron-rich Kr}
\begin{figure}[tb]
\centering\includegraphics[width=0.95\textwidth]{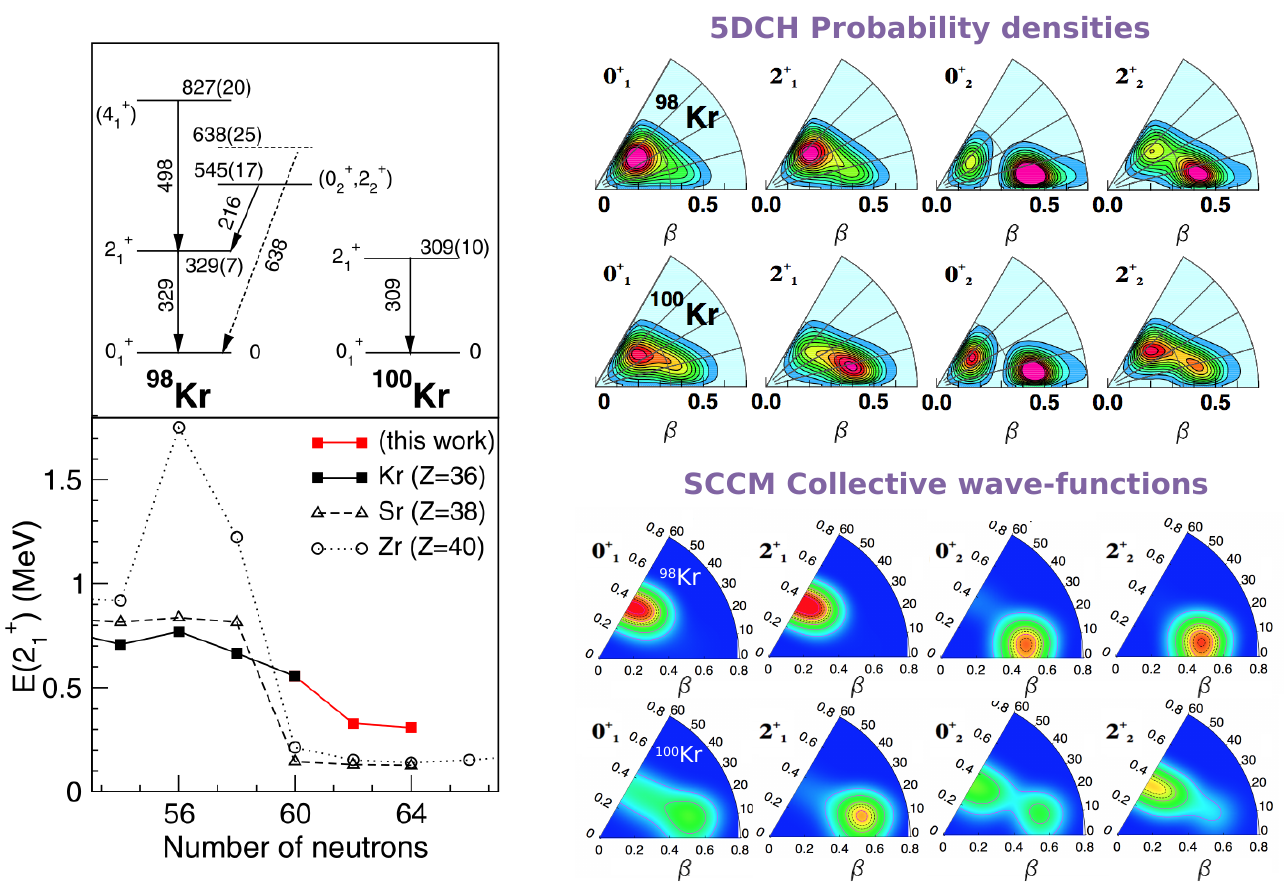}
\caption{Excitation states of \ts{98,100}Kr isotopes, together with the probability densities from 5DCH calculations -with a metric~\cite{Delaroche_2010_PRC}- and collective wave functions from SCCM calculations for the yrast and nonyrast bands. The figure is adapted from Ref.~\cite{Flavigny_2017_PRL}. Reprinted figure with permission from Ref.~\cite{Flavigny_2017_PRL}. Copyright 2017 by the American Physical Society.}
\label{fig:Kr_level_shape}
\end{figure}

The neutron-rich Kr isotopes are located close to the rapid shape transitions $A\simeq100$ region at $N=60$ for Zr ($Z=40$) and Sr ($Z=38$) isotopes~\cite{Heyde_2011_RMP}, and were recognized as the critical-point boundary of this region~\cite{Naimi_2010_PRL}.
A picture of gradual development of collectivity in Kr isotopes up to $N=60$ was well-established from the charge radii~\cite{Keim_1995_NPA}, the two-neutron separation energies $S_{2n}$~\cite{Naimi_2010_PRL}, the smooth reduction of $E(2^+_1)$ and the rise of $B(E2,0^+_1\rightarrow2^+_1)$ excitation strength~\cite{Albers_2012_PRL}.
However, it cannot exclude a shape transition in the unexplored $N>60$ region.
In this campaign, the first spectroscopy of \ts{98,100}Kr isotopes were obtained following \ts{99,101}Rb($p$,$2p$) reactions.
The obtained level schemes and the systematics of the $E(2^+_1)$ are shown in Fig.~\ref{fig:Kr_level_shape}.
A marked decrease of the $2^+_1$ state energy at $N=62$ demonstrates a significant increase of deformation.
In \ts{98}Kr, a ($0^+_2,2^+_2$) state close to the $2^+_1$ state was identified, providing the first experimental evidence for a coexisting band.
The results were compared with two beyond-mean-field calculations using the Gogny D1S interaction: the five-dimensional collective Hamiltonian (5DCH) and the SCCM, both of which well reproduce the experimental energy levels including the low-lying ($0^+_2,2^+_2$) state.
The probability densities from 5DCH and the collective wave functions from SCCM are also presented in Fig.~\ref{fig:Kr_level_shape}, revealing similar prolate-oblate shape-coexistence and transition pattern.
These observations highlight the rich collective behaviours in this $N\simeq60, A\simeq100$ shape transition region. It is important to point out that 5DCH wave functions include a metric in the $(\beta_{2},\gamma)$-plane that produces probability densities equal to zero at the axial-symmetric configurations, i.e., $\beta_{2}=0$ and $\gamma=0^{\circ},60^{\circ}$~\cite{Delaroche_2010_PRC}. 

The $\gamma$-ray spectroscopy of low-lying excited states in neutron-rich \ts{94,95,96}Kr isotopes were also measured in the same experiment~\cite{Gerst_2022_PRC}.
For the even-even \ts{94,96}Kr isotopes, transitions from excited yrast and non-yrast states were observed following the nucleon removal reactions. The results confirmed the observations reported in Ref.~\cite{Dudouet_2017_PRL,Gerst_2020_PRC}, and significantly extended the level scheme for \ts{96}Kr.
The candidates for $3^-_1\rightarrow2^+_1$ transitions were identified from the $(p,p')$ inelastic scattering reactions and indicate the octupole collectivities in the neutron-rich Kr isotopes.
The experimental results were compared with 5DCH and mapped interacting boson model (IBM) calculations, both using the Gogny D1M interaction.
The 5DCH calculations provides a good description of the excitation states and transition strengths, and suggest the prolate-oblate shape coexistence already presented in \ts{96}Kr.
For \ts{95}Kr, the in-beam $\gamma$-ray spectroscopy measurements were performed in conjunction with decay spectroscopy using the EURICA array. Through measuring the in-beam spectrum measured in DALI2 in coincidence with the $\gamma$ rays from the decay of the known $(7/2)^+$ isomer~\cite{Genevey_2006_PRC}, the structure above it could be investigated.
 Three transitions were tentatively placed on top of the isomeric decay using the delayed-prompt coincidences. Future measurements with high resolution spectroscopy will be necessary to build a level scheme. 

\subsection{Deformation maintained at the harmonic oscillator magic numbers in \ts{110}Zr}
\begin{figure}[tb]
\centering\includegraphics[width=0.85\textwidth]{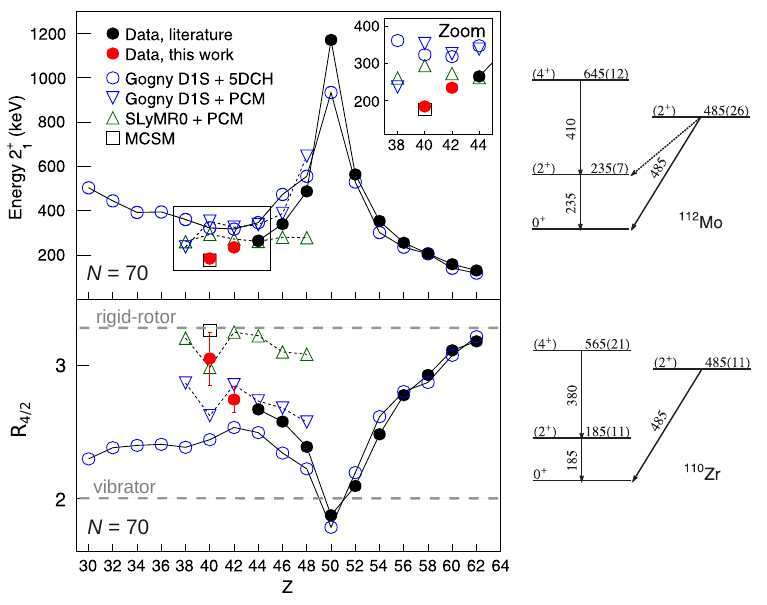}
\caption{Systematics of $E(2^+_1)$ and $R_{4/2}$ ratio for the $N=70$ isotones. The grey dashed lines show the vibrator limit, 2, and rigid-rotor limit, 3.3, for $R_{4/2}$ ratio. The figure is taken from Ref.~\cite{Paul_2017_PRL}. Reprinted figure with permission from Ref.~\cite{Paul_2017_PRL}. Copyright 2017 by the American Physical Society.}
\label{fig:N70_level_shape}
\end{figure}

The \ts{110}Zr nucleus has 40 protons and 70 neutrons, both corresponding to harmonic oscillator magic numbers. If the spin-orbit splitting that gives rise to the $N=82$ shell gap is reduced in the nuclei far from stability, the harmonic oscillator gap at $N=70$ may open up, leading to a shell-stabilized \ts{110}Zr.
Such shell-stabilization in \ts{110}Zr has been predicted by several independent mean-field and microscopic-macroscopic approaches~\cite{Dudek_2002_PRL,Schunck_2004_PRC,Zhao_2017_PRC}, while the other theoretical calculations predicted a well-deformed shape or shape coexistence in \ts{110}Zr~\cite{Geng_2003_PTP,Kortelainen_2010_PRC,Delaroche_2010_PRC,Skalski_1997_NPA,Xu_2002_PRC,Shi_2012_PRC,Petrovici_2011_JPCS,Togashi_2016_PRL}.
Spectroscopic evidence of the deformed nature of \ts{110}Zr and \ts{112}Mo was observed in this campaign. Figure~\ref{fig:N70_level_shape} shows the measured level schemes, and the systematics of $E(2^+_1)$ and $R_{4/2}$ ratio for the $N=70$ isotones.
As one departs from the $Z=50$ shell closure, the $E(2^+_1)$ decreases and the $R_{4/2}$ ratio increases from below 2 to 3.1(2) at \ts{110}Zr. This pattern is consistent with a transition from shell closure to midshell nuclei, i.e., from a harmonic vibrator to a deformed rotor.
The systematics down to $Z=42$ are overall well reproduced by the 5DCH and PCM calculations using the Gogny D1S interaction. However, both calculations over predict the energy of the $2^+_1$ level in \ts{110}Zr, which is interpreted as an under prediction of its collectivity. Results of the MCSM calculations, which are only performed for \ts{110}Zr, agree well with the measured values. 

\section{Perspectives for high-resolution $\gamma$-ray spectroscopy}
Key challenges of the in-flight spectroscopy of deformed exotic nuclei arise through their excited states being relatively low in excitation energy and rather long-lived. The first of these poses a challenge since there is a substantial background at low energies associated with Brehmstrahlung, which becomes prominent in spectra below around 200-300~keV. Particular care had to be taken of this component in the spectroscopy of \ts{110}Zr and \ts{112}Mo~\cite{Paul_2017_PRL}. The long lifetimes of low-lying deformed states can introduces a systematic lowering of the measured $\gamma$-ray energies, however, they can prove a a powerful tool in the determination of transition probabilities, discussed below. The spread of the decay position due to lifetime effects and the velocity of the $\gamma$-ray emission particle introduces a large uncertainty to the $\gamma$-ray emission angle, thereby degrading the Doppler-reconstructed energy. 

Distinguishing $\gamma$ rays in the low-energy background can be partly mitigated by high-resolution $\gamma$-ray spectrometers, where sharper peaks are more easily distinguished from background contributions. As alluded to previously, the degradation of the Doppler-corrected peaks from lifetime effects can be used to evaluate the lifetimes through the comparison to the simulated response functions, provided a sufficiently high-resolution $\gamma$-ray spectrometer is utilised~\cite{Alexander1978}. In principle, the peak becomes ``smeared'' to lower energies. The degree of this smearing can be translated into lifetimes of excited states through comparison to sufficiently accurate simulations of the $\gamma$-ray spectrometer. As well as these advantages, the enhanced resolving power also allows for the clear separation of closely-spaced transitions, particularly important for the cases described above where the $2^+_1$ and $2^+_2$ states become closer in energy. 

Efforts to introduce high-resolution spectroscopy at the RIBF recently culminated in the execution of the HiCARI campaign~\cite{wimmer_rapr_2021}, which utilised a mixed array of high-purity germanium detectors for in-beam spectroscopy. The intrinsic resolution of the HiCARI array compared to DALI, used for all the results presented previously, overcomes the limitations described above. Relevant to the current proceedings, experiments aimed at the study of transition probabilities in \ts{86,88,90}Se and \ts{84,86}Ge~\cite{browne_rapr_2021}, intruder states in Zn isotopes~\cite{flavigny_rapr_2021}, and measuring lifetimes in neutron-rich Zr/Mo isotopes~\cite{korten_rapr_2021,moon_plb_2024} were performed. 

Ongoing analysis of the HiCARI experiments listed above have demonstrated the advantages of high-resolution spectroscopy. In many cases, it appears that lifetimes can be extracted from the measurements. For Ref.~\cite{browne_rapr_2021}, the additional sensitivity to weakly-populated states provides the ability to measure Coulomb excitation cross sections to non-yrast states. From knockout reactions, Ref.~\cite{korten_rapr_2021,moon_plb_2024} has revealed a more detailed level scheme of \ts{110}Zr, as well as providing measurement of many excited-state lifetimes. With many interesting results now in preparation, the results from the HiCARI campaign will prove complimentary to the cutting-edge work performed with the MINOS device coupled with the DALI2 spectrometer. 

It is prudent to outline the desirability of high-resolution decay spectroscopy of stopped nuclei within the context of the work discussed here. The measured energies of $\gamma$ rays emitted from rest are insensitive to the lifetimes effects discussed above. Decay measurements also provide a greater sensitivity to low-intensity transitions that help with building comprehensive level schemes, including accurately determining branching ratios of states whose lifetimes can be measured through in-beam $\gamma$-ray spectroscopy. Such measurements are important for the determination of the reduced transition probabilities which can be compared to theoretical predictions. In addition to all this spectroscopic information, the usefulness of isomer-tagging of in-beam spectra from decay $\gamma$ rays is outlined above, this capability can be further exploited by ensuring more efficient arrays are developed, especially for higher-energy $\gamma$ rays.

The major challenge associated with decay spectroscopy measurements is that of RI beam production. With knockout reactions, production of isotopes less exotic than the isotope of interest are required, whilst decay studies require the production of more exotic isotopes. The capabilities of new facilities such as FRIB and upgrades to the RIBF strive towards the production of extremely neutron-rich nuclei in sufficient abundance to perform the decay studies on those nuclei described here, so far only broached through spectroscopy following direct reactions. 

\section{Conclusion}
The deployment of LH$_2$ targets at the RIBF in conjunction with the high-efficiency DALI2 $\gamma$-ray spectrometer has greatly expanded the spectroscopic information of nuclei across the $N\approx60$ shape transitional region. The 2\ts{nd} SEASTAR campaign provided the first spectroscopy of \ts{84}Zn~\cite{shand_plb_2017}, \ts{88}Ge~\cite{PRC_95_041302_2017}, \ts{90,92,94}Se~\cite{PRC_95_041302_2017}, \ts{98,100}Kr~\cite{Flavigny_2017_PRL}, \ts{110}Zr, and \ts{112}Mo~\cite{Paul_2017_PRL}, as well as the expansion of level schemes of many other nuclei. These results have charted the evolution of deformation deep into the doubly mid-shell region between $28<Z<50$ and $50<N<82$, demonstrating no stabilisation of the spherical shape past the $N\approx60$ shape transition.

Interpretation of the experimental results requires an approach in which intrinsic shapes are well-defined and arise naturally from the model. The PGCM provides such a framework, and has been shown to reproduce the excitation spectra of in the Ge, Se and Kr isotopic chains, crucially including the $\gamma$ bands, which reflect triaxiality, and shape coexistence. 

Partially motivated by the results described here, the HiCARI campaign was conducted utilising high-resolution $\gamma$-ray spectroscopy to provide further detail to the novel deformed structures that typify the neutron-rich $28<Z<50$. A notable difference to this work was the use of compound targets, as well as a mixed array of detectors with differing experimental responses. Future experimental works may involve using an array of HPGe detectors that can track multiple $\gamma$-ray scatterings inside a crystal to determine its emission angle. Such a tracking detector array combined with a MINOS-like target to exploit direct reactions would achieve a superiour in-beam $\gamma$-ray energy resolution. Such developments at the RIBF are detailed in Ref.~\cite{S25V-008}.

\section*{Acknowledgment}

SC acknowledges support from the UK STFC and the Royal Society. 
TRR acknowledges the Grant PID2021-127890NB- I00 funded by MCIN/AEI/10.13039/501100011033, and the support of the GSI-Darmstadt computer facilities.
This article is based on work presented at the symposium ``Direct Reactions and Spectroscopy with Hydrogen Targets: Past 10 years at the RIBF and Future Prospects'', 31st July to 4th August 2023, and supported by the IOP and the JSPS London Symposium and Seminar Scheme (award number: JBUK028).


\raggedbottom

\bibliographystyle{ptephy}
\bibliography{bibliography}
%





\end{document}